\documentclass[11pt,letterpaper,twocolumn]{article}
\usepackage[utf8]{inputenc}
\usepackage{graphicx}
\usepackage{hyperref}

\fussy
\begin{document}

\begin{center}
{\bf Extending the NOvA Physics Program} \\
{\it Mark Messier, for the NOvA Collaboration }
\end{center}

\noindent
Following the precise measurement of $\theta_{13}$ by reactor experiments~
\cite{Abe:2011fz,An:2012eh,Ahn:2012nd,An:2012bu} 
three main questions remain within the now standard picture of neutrino oscillations: 
(1) Is the value of $\theta_{23}$ such that the $\nu_3$ state contains more muon flavor, more tau flavor, or equal amounts?
(2) What is the neutrino mass hierarchy? 
(3) Is CP violated in the neutrino sector?
The NOvA experiment~\cite{Ayres:2007tu} will address all three of these questions. In each case the measurements are statistics limited motivating exploration of what could be accomplished with additional exposure.

\begin{figure}[h]
\begin{center}
\includegraphics[width=2.75in]{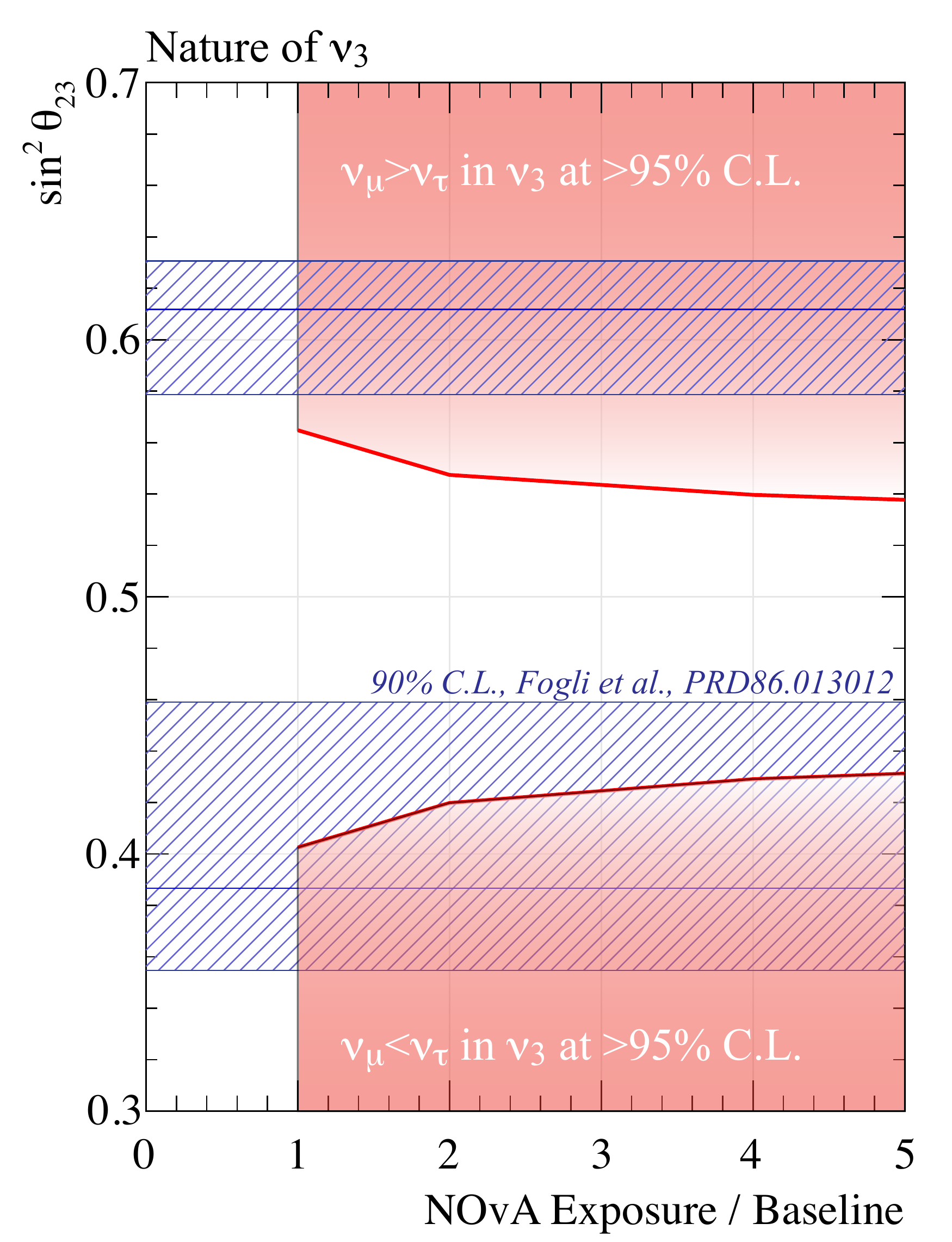}
\caption{NOvA's reach to resolve the nature of $\nu_3$ as a function of the experiment exposure. For maximal mixing ($\sin^2 \theta_{23}=0.5$) $|\langle \nu_3 | \nu_\mu \rangle|^2 =
|\langle \nu_3 | \nu_\tau \rangle|^2$ with the muon content of $\nu_3$ exceeding the $\tau$ content for $\sin^2 \theta_{23}>0.5$. NOvA can determine the relative sizes of  
$|\langle \nu_3 | \nu_\mu \rangle|^2$ and
$|\langle \nu_3 | \nu_\tau \rangle|^2$ at 95\% C.L. for the values of $\sin^2 \theta_{23}$ shaded in red. The blue hatched region indicates current world knowledge of the possible values of $\sin^2 \theta_{23}$.}
\label{fig:nova-coverage-octant}
\end{center}
\end{figure}

The baseline exposure for the NOvA experiment assumes a 14~kt detector, 700~kW NuMI beam power, and 6 years of running. Doubling the NOvA exposure would require a relatively modest investment, have low risks, and would leverage the substantial investments made in the NuMI beam, the Ash River laboratory site, and the setup of the NOvA production factories. Construction of the NOvA detector is underway; 1/4 of the detector is in place and is being filled with scintillator. 
An increase in exposure of 2.1$\times$ can be realized by increasing the detector mass to 18~kt and extending the run to 10~years. This run plan uses all of the available space in the laboratory and recognizes the schedule realities of next-generation projects. If construction of the NOvA detector is continued without interruption, the cost of additional mass would be \$6M/kt. A conservative upper limit which assumes that current construction ends and that all the start-up costs must be paid again on future construction raises this figure to \$9M/kt.

\begin{figure}[t]
\begin{center}
\includegraphics[width=2.75in]{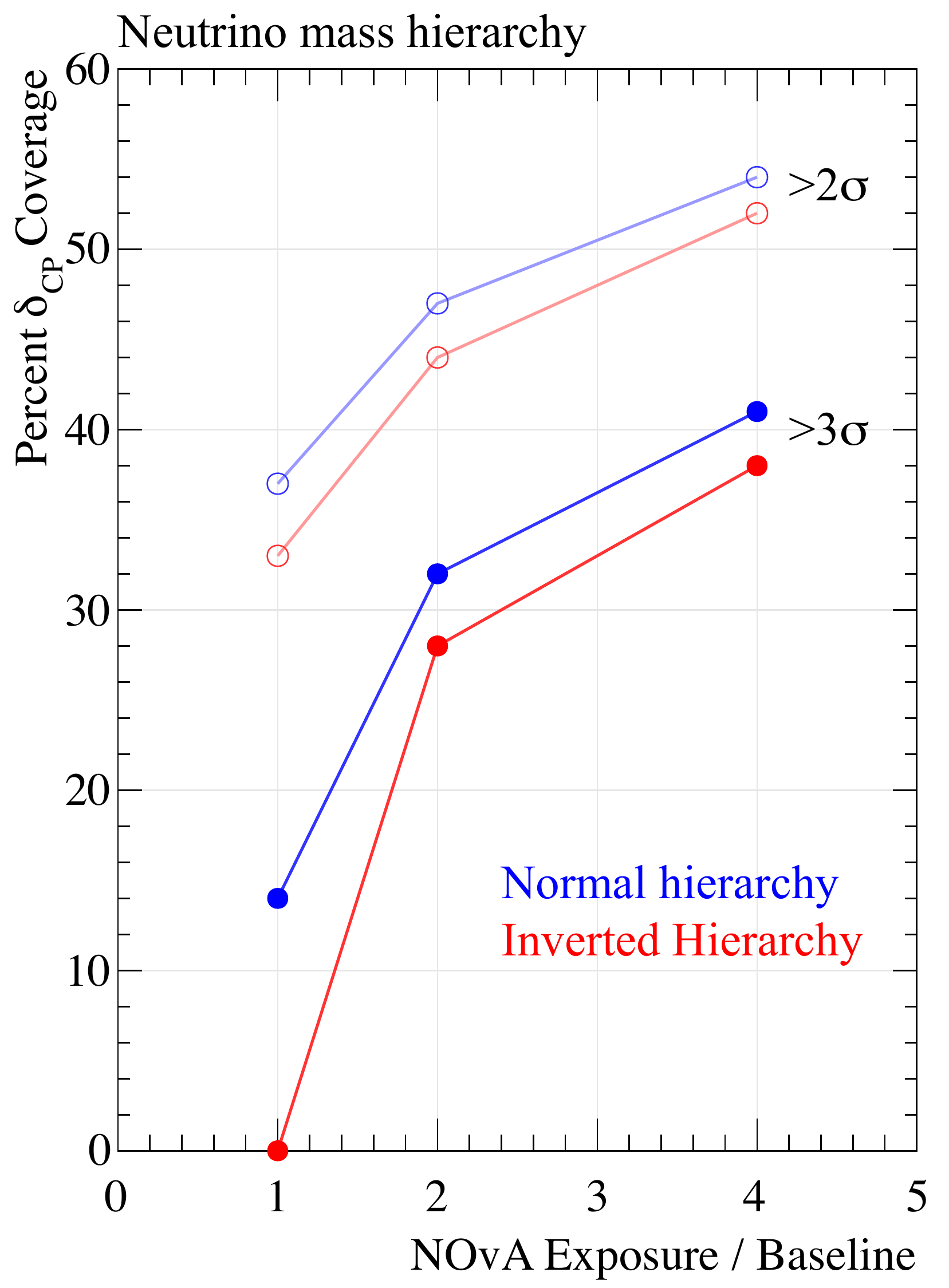}
\caption{The percent of $\delta_{\rm CP}$ values for which NOvA can resolve the neutrino mass hierarchy at 2 and 3$\sigma$~C.L.}
\label{fig:nova-coverage-hier}
\end{center}
\end{figure}

Additional exposure could follow in a second phase and is motivated by the LBNE reconfiguration report~\cite{LBNERecon} which found Ash River to be the site with maximum CP reach assuming that the mass hierarchy is resolved by the experiments planned for this decade (eg. NOvA, Pingu, Daya Bay II). A 5~kt liquid argon TPC at the Ash River site, either in the NOvA laboratory or in a new facility which reuses the infrastructure supporting the NOvA laboratory, effectively increases the NOvA exposure by a factor of 4 given the improved performance of liquid argon detectors.

\begin{figure}[t]
\begin{center}
\includegraphics[width=2.75in]{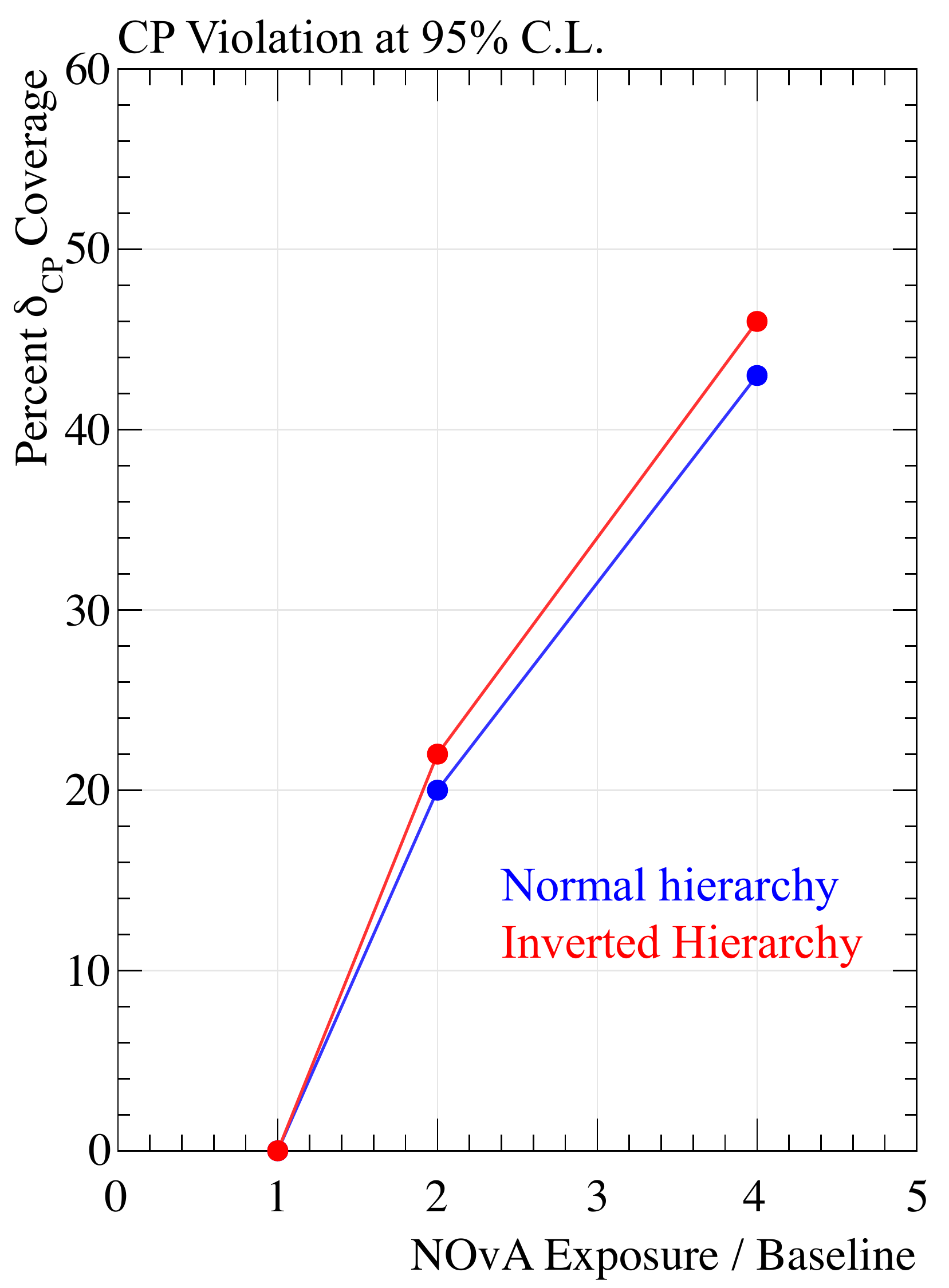}
\caption{The percent of $\delta_{\rm CP}$ values for which NOvA can establish CP violation at 95\% C.L. or better.}
\label{fig:nova-coverage-cp}
\end{center}
\end{figure}

Figures~1-3 outline what is possible with additional exposure. Figure~\ref{fig:nova-coverage-octant} shows the extended reach for resolving the nature of $\nu_3$ relative to the current knowledge of $\sin^2 \theta_{23}$ following Neutrino 2012. NOvA's baseline measurement covers 64\% of the currently allowed 90\% C.L. region at 95\% C.L. or better. With 2$\times$ the exposure this increases to 75\% and 80\% for 4$\times$. Figure~\ref{fig:nova-coverage-hier} shows the improvement in mass hierarchy resolution. With additional exposure, a significant amount of coverage is obtained at $>3~\sigma$ over the baseline experiment. Finally, NOvA's reach for CP violation increases rapidly with exposure in Figure~\ref{fig:nova-coverage-cp}. NOvA's baseline exposure enables a first measurement of $\delta_{\rm CP}$ but the precision will not be enough to establish CP violation. CP violation can be established with 95\% C.L. for 20\% of the $\delta_{\rm CP}$ space for $2\times$ the exposure, increasing to 45\% for $4\times$ the exposure.

In summary, a modest investment to extend the NOvA exposure to 2$\times$ its baseline through a combination of detector mass and running time would yield qualitative improvements in the experiment's hierarchy and CP violation reach. A 5~kt liquid argon TPC at the Ash River site could extend the physics reach further in a second phase. These extensions would leverage the investments made in the NOvA factories, the Ash River laboratory, and the NuMI beam.

\end{document}